\documentclass[twocolumn,showpacs,preprintnumbers,amsmath,amssymb,prb]{revtex4-1}

\usepackage[active]{srcltx}
\usepackage{graphicx}
\usepackage{dcolumn}
\usepackage{bm}

\newcommand{\ii}{{\rm i}}

\def\gsim{\lower.35em\hbox{$\stackrel{\textstyle>}{\textstyle\sim}$}}
\def\lsim{\lower.35em\hbox{$\stackrel{\textstyle<}{\textstyle\sim}$}}

\begin{document}

\title{Fluorescence quenching in graphene: a fundamental ruler and evidence for transverse plasmons}

\author{G. G\'omez-Santos and T. Stauber}

\affiliation{Departamento de F\'{\i}sica de la Materia Condensada  and Instituto Nicol\'as Cabrera, Universidad Aut\'onoma de Madrid, E-28049 Madrid, Spain}

\begin{abstract}
Graphene's fluorescence quenching is studied as a function of
distance.  Transverse decay channels, full retardation and
graphene-field coupling to all orders are included, extending previous
instantaneous results. For neutral graphene, a virtually exact
analytical expression for the fluorescence yield is derived, valid for
arbitrary distances and only based on the fine structure constant
$\alpha$, the fluorescent wavelength $\lambda$, and distance $z$. Thus
graphene's fluorescence quenching measurements provide a fundamental
distance ruler. For doped graphene and at appropriate energies, the
fluorescence yield at large distances is dominated by transverse
plasmons, providing a platform for their detection.
\end{abstract}


\maketitle


\section{Introduction}
The optical properties of graphene have
attracted immense attention due to the potential applications in the
field of photonics and optoelectonics \cite{Bonaccorso10}. Absorption,
for instance, has the universal value $\approx \pi \alpha$ for light
in the visible spectrum, depending on the fine structure constant
$\alpha$, but not on material's properties\cite{Nair08,Mak08},
providing a decisive clue for its
original identification\cite{Blake07}. The large intrinsic carrier
mobilities and doping tunability have led to a number of proposals for
optoelectronic
applications\cite{Vakil11,Bonaccorso10,Xia09,Avouris10}, where the
engineering of long-lived graphene's plasmons could play a major
role\cite{Jablan09,Koppens11,Nikitin11}.

Graphene (and its oxide) exhibits excellent quenching of nearby
fluorescent materials\cite{Treossi09,Kim10,Chen10,Loh10}, a property
shared with carbon nanotubes\cite{Qu02,Nakayama07}. This technique has
allowed spectacular contrast images, enabling far easier optical
identification\cite{Kim10} (and prospects for device
manipulation\cite{Sagar10}) of graphene's flakes. Given the mature
nature of fluorescent microscopy, particularly in the biological
sciences, their combination with increasingly available and versatile
graphene nanostructures could open interesting research
directions\cite{Loh10}. Furthermore, graphene's fluorescence quenching
has also been proposed as a convenient probe of the intrinsic
excitations of doped graphene such as plasmons\cite{Velizhanin11}.

Here, we theoretically address the effect of graphene on the
fluorescent material's yield as function of the distance $z$ within a
unifying formalism. By this, we discuss for the first time the
transverse decay channels known to lead to characteristic features
only found in graphene.\cite{Ziegler07} The process implies
non-radiative (F\"{o}rster\cite{Forster48,Chance78}) decay of the
excited dye, with energy transfer to graphene's
excitations. This mechanism is expected to dominate over competing
charge-transfer processes except, perhaps, in near contact
situations\cite{Lin10,Klekachev11}. Our motivation stems from the
seemingly paradoxical coexistence in graphene of very strong
fluorescence quenching and nominally weak coupling to the
electromagnetic field, as judged from the absorption results.

The distance behavior of fluorescence quenching as function of the
distance $z$ of a dye molecule with respect to a metal surface was
found long ago to be $z^{-4}$.\cite{Persson82} This law has 
repeatedly been found in subsequent studies of energy transfer involving
various kinds of nanomaterials.\cite{Yun05} For graphene, the topic
has been discussed by Swathi and Sebastian, treating the
(instantaneous) longitudinal coupling between graphene and fluorescent
material to lowest order, and again obtaining the
$z^{-4}$-law.\cite{Swathi09} Here, we extend their calculations in
three significant ways: i) we include transverse decay channels in the
calculations ii) the coupling between graphene and the electromagnetic
field is taken to all orders iii) we use the full (retarded) photon
propagator.

Our results for undoped graphene are as follows: i) a compact,
analytical approximation is obtained and shown to provide a virtually
exact description of the fluorescent yield for all distances.  This
analytical expression only depends on the ratio of the distance to the
emitting wavelength $(z/\lambda)$ and the fine structure constant
$\alpha$. A fluorescence measurement thus becomes a distance ruler,
based only on fundamental constants - a long sought goal of the field. ii)
Retardation is shown to modify the $z^{-4}$ law into a slower $z^{-2}$
behavior, with the transverse decay channel dominating at long
distances.

For doped graphene, we show that the fluorescence yield is mostly
determined by the plasmonic modes\cite{Velizhanin11} where at certain
frequencies transverse plasmons yield the dominant, extremely slowly
decaying contribution at large distances. This should help to detect
these modes characteristic to graphene and intimately linked to the
chirality of its elementary excitations.

The paper is organized as follows. In section II, we introduce the formalism defining the atom-field coupling in the presence of graphene and the induced fluorescence quenching. In section III, we present our results and then close with a summary. In an appendix, details on the explicit form of the decay rates are given.

\section{Formalism} 
\subsection{Atom-field coupling in the presence of graphene}
We consider a fluorescent atom modeled by a two level system and the
electromagnetic field described in a gauge without scalar potential. 
%
%
%
%
%
%
%
%
%
Within the standard dipole approximation, 
an excited atom will decay into the ground state at a rate $(\gamma)$ 
given by\cite{Novotny06}
\begin{equation}\label{gamma}
\hbar \gamma = -2 \omega^2 p_{\alpha}^{*}  \;\text{Im} 
             \mathcal{D}^{\alpha \beta}(\bm r,\bm r;\omega) \;p_{\beta}
,\end{equation} 
where $\bm p$ 
is the dipole matrix
element between ground and excited states,
 separated in energy by $ \hbar \omega$. $\mathcal{D}$ represents 
 the retarded photon Green's function defined as usual,
\begin{equation}\label{Ddefinition}
 \mathcal{D}^{\alpha \beta}(\bm r,\bm r';\omega) = -\frac{\ii}{\hbar}
 \int_{0}^{\infty} dt \,{\rm e}^{\ii\omega t}\, 
 \langle [ A_{\alpha}(\bm r,t), A_{\beta}(\bm r',0)]\rangle
,\end{equation} 
where $\bm A(\bm r)$ is the vector potential operator at the atom's location.
Notice that Eq. \eqref{Ddefinition} corresponds to the photon propagator without
 the fluorescent atom, but in the presence of graphene.

 In vacuum (that is, {\em without} graphene),  the photon Green's function
$\mathcal{D}_0^{\alpha \beta}(\bm r-\bm r';\omega)$ is given by the Fourier
transform of
$\mathcal{D}_0^{\alpha \beta}(\bm k,\omega) =
 \tfrac{\mu_0 c^2}{\omega^2-c^2 k^2} (\delta_{\alpha \beta} -
 \tfrac{k_{\alpha} k_{\beta}}{k^2}) +\tfrac{1}{2 \epsilon_0 \omega^2} 
 \tfrac{k_{\alpha} k_{\beta}}{k^2}$. 
%
%
Later inclusion of graphene, assumed perpendicular to the $z$ axis, will
preserve the parallel (to graphene's plane) component of momentum, 
$\bm q = (q_1,q_2)$, 
as a good quantum number. 
Therefore, it is convenient to employ the following representation for the 
vacuum Green's function
\begin{equation}\label{D0zz'}
 \mathcal{D}_0^{\alpha \beta}(z,z';\bm q,\omega) = \frac{1}{2\pi}
 \int dk_z \, {\rm e}^{\ii k_z (z-z')} \, 
 \mathcal{D}_0^{\alpha \beta}(\bm k,\omega)
,\end{equation} 
with $\bm k = (\bm q,k_z)$. Physically, Eq. \eqref{D0zz'} represents the
vector
potential in a plane perpendicular to the z-axis located at the position $z$
due to currents in a parallel plane at location $z'$.


The in-plane components of the tensor $ \mathcal{D}_0^{ij}$, 
decomposed into longitudinal and transverse contributions, are
given by
\begin{equation}\label{D0explicit}
 \mathcal{D}_0^{i j}(z,z') = d_{l} {\rm e}^{-q'|z-z'|}\,
                                          \frac{q_i q_j}{q^2} + 
					  d_{t} {\rm e}^{-q'|z-z'|}\, 
		(\delta_{i j}-\frac{q_i q_j}{q^2})   
,\end{equation} 
with $i(j) = 1,2 $ and  $q' = \sqrt{q^2-(\omega/c)^2}$. The functions 
$d_{l,t}(\bm q,\omega)$ are given by (dependencies removed for clarity)
\begin{equation}\label{dlt}
d_l = \frac{q'}{2 \epsilon_0 \omega^2}, \;\;\;\;
d_t = -\frac{c^{-2}}{2 \epsilon_0 q'}
.\end{equation} 
The remaining tensor components are written as
\begin{equation}\label{D0izexplicit}
 \mathcal{D}_0^{i z}(z,z') = \mathcal{D}_0^{z i}(z,z')=
  \frac{\ii q_i}{q'}d_l {\rm e}^{-q'|z-z'|} \,{\rm sgn}(z-z')  
\end{equation} 
 and 
%
\begin{equation}\label{D0zzexplicit}
 \mathcal{D}_0^{z z}(z,z') =
  \frac{1}{\epsilon_0 \omega^2} \delta(z-z') - 
   \frac{q^2}{q'^2}d_l {\rm e}^{-q'|z-z'|} 
.\end{equation} 

The presence of a graphene plane at the location $z_1$ modifies the vacuum
Green's function as follows 
\begin{equation}\label{Dgraphene}
 \mathcal{D}^{\alpha \beta}(z,z') =
 \mathcal{D}_0^{\alpha \beta}(z,z') + 
 \mathcal{D}_0^{\alpha i}(z,z_1)
 e^2\chi^{ij}
 \mathcal{D}_0^{j \beta}(z_1,z')
,\end{equation} 
where sum over repeated indexes is assumed. $ e $ is the  electron charge  and $\chi^{i j}(\bm q,\omega)$ represents 
graphene's current-current total response to
external fields. The latter, decomposed into longitudinal and transverse
contributions, is  given by
\begin{equation}\label{chiij}
\chi^{i j} = \frac{\chi^{}_l}{1 - e^2 d_l \chi^{}_l} \frac{q_i q_j}{q^2} + 
             \frac{\chi^{}_t}{1 - e^2 d_t \chi^{}_t} (\delta_{i j}-\frac{q_i q_j}{q^2})
,\end{equation} 
where we take the non-interacting (RPA), well-known expression for the
longitudinal\cite{Shung86} and transverse\cite{Principi09} components at zero doping:  
%
%
%
\begin{equation}\label{chilt}
\chi^{}_l = -\frac{g_s g_v}{16 \hbar v} \frac{\omega^2}{\sqrt{q^2-(\omega/v)^2}}\;
,
\;\;\;\;
\chi^{}_t = \frac{g_s g_v}{16 \hbar} v \sqrt{q^2-(\omega/v)^2}
,\end{equation} 
with spin and valley degeneracies, $ g_s=g_v=2$, and graphene's velocity $v$. For finite doping, we refer to the expressions given in Refs. \cite{Principi09,Stauber10}.

The previous calculation of Swathi and Sebastian\cite{Swathi09} would correspond
to zero doping and retaining only the (numerator of the) longitudinal response $(\chi^{}_l)$ in
Eq. \eqref{chiij}, while  setting $c\rightarrow
\infty$ in the photon propagator (instantaneous limit).

\subsection{Fluorescence quenching} 
Consider the graphene sheet placed at the origin ($z_1=0$) and the excited atom at a distance $z$. The expression \eqref{gamma} for the decay rate can be decomposed as
\begin{equation}\label{gammarnr}
\hbar \gamma^{}_{\stackrel{nr}{r}} = -2 \omega^2 p_{\alpha}^{*}  \; 
\{\frac{1}{(2 \pi)^2}\int_{q\gtrless\omega/c} d^2q \,
                   \text{Im}\mathcal{D}^{\alpha \beta}(z,z)\}             
                \;p_{\beta}\;,
\end{equation} 
with the (graphene's modified) photon Green's function given by
Eq. \eqref{Dgraphene}. The $\bm q$ label  classifies the final field states into
{\em evanescent} excitations $(q > \omega/c)$, and  {\em propagating} 
excitations $(q < \omega/c)$, the latter being the observed photons.
Therefore, the total decay rate is given by the radiative and non-radiative contributions to the decay rate, 
\begin{equation}\label{gammarplusnr}
\gamma = \gamma^{}_{nr} + \gamma^{}_r\;.
\end{equation} 
Let us consider  the rate of {\em observed} photons $\Phi$. In addition to 
$\gamma^{}_{r,nr}$,
 it will depend on the
rate at which the atom is pumped into the excited state $\gamma^{}_{exc}$. 
Furthermore, not all
propagating photons are observed, a fraction being later absorbed by 
graphene $\gamma^{}_{abs}$.
%
%
%
The excitation rate is hardly affected by the presence of
graphene and the fraction of emitted photons later absorbed is, up to
logarithmic corrections, of the order of the fine structure constant 
$(\alpha=\frac{e^2}{4 \pi \epsilon_0 \hbar c})$.
Therefore, the ratio of the total observed fluorescence when the atom is at distance
$z$, $\Phi(z)$, to that at infinite distance, $\Phi_{\infty}$, can be written as 
\begin{equation}\label{PhiPhi}
\frac{\Phi(z)}{\Phi_{\infty}} = \left(1 +\frac{\gamma^{}_{nr}}{\gamma^{}_{r}}\right)^{-1}
,\end{equation} 
where the  neglected terms  amount to minute relative corrections of order $\alpha^2$ in the expression \eqref{PhiPhi}.

\section{Results}
\subsection{Zero doping}
\label{results} 
 We have evaluated the distance dependence to undoped graphene
of the observed fluorescence, Eq. \eqref{PhiPhi}, with $\gamma^{}_{r,nr}$ obtained from Eq.
\eqref{gammarnr} and Eq. \eqref{Dgraphene}.  There is a sharp difference in graphene's effect
on the non-radiative and radiative contributions to the decay. Graphene modification of the
(radiative) vacuum decay is a weak effect, proportional to $\alpha$ (up to logarithmic
corrections). Therefore, setting $\gamma^{}_{r} \approx \gamma^{}_0$, with the
vacuum decay rate given by $\hbar\gamma^{}_0 = \frac{p^2 \omega^3}{3\pi \epsilon_0 c^3}$, 
the results can be written as
%
\begin{equation}\label{gammaanal}
\frac{\gamma^{}_{nr}}{\gamma^{}_{r}} \approx \frac{\gamma^{}_{nr}}{\gamma^{}_{0}} = 
\beta_1 \tilde{\gamma}^{}_1 \tilde{f}^{}_1 + \beta_2 \tilde{\gamma}^{}_2 \tilde{f}^{}_2+ \beta_3\tilde{\gamma}^{}_3\tilde{f}^{}_3  
,\end{equation} 
with the physically relevant magnitudes given by 
\begin{subequations}\label{items}
\begin{align}
\tilde{\gamma}^{}_1= \;&\frac{3^2}{2^9\pi^3} \; \alpha \;
\left(\frac{\lambda}{z}\right)^4 \label{first}
\\\tilde{\gamma}^{}_2= \;&\frac{3}{2^6\pi} \; \alpha \;
\left(\frac{\lambda}{z}\right)^2 \label{second}  
\\\tilde{\gamma}^{}_3= \;&\frac{3}{4}\pi \; \alpha \; {\rm g}(2 \pi^2
\alpha z/\lambda) 
\xrightarrow
[\frac{z}{\lambda} \gtrsim \frac{1}{2 \pi^2 \alpha}]{}
\frac{3}{2^4\pi^3\alpha}\left(\frac{\lambda}{z}\right)^2    
\label{third}
,\end{align}             
\end{subequations} 
where the function ${\rm g}(a)$ can be written in terms of the sine and 
cosine-integrals (${\rm si},{\rm ci}$)
 as
\begin{equation}\label{g(a)1}
{\rm g}(a) = -{\rm ci}(a) \cos(a) - {\rm si}(a)\sin(a)
.\end{equation}

The coefficients $\beta_{1,2,3}$ are mere geometric
factors depending on the emitting dipole orientation, with 
$\beta_1=(p_{\shortparallel}^2/2 + p_z^2)/p^2$, $\beta_2=p_z^2/p^2$, and 
$\beta_3=(p_{\shortparallel}^2/2)/p^2$. 
All information about graphene in Eq. \eqref{gammaanal} is relegated to the dimesionless 
factors $\tilde{f}_{i}$, derived and discussed in the appendix.

The first term, $\tilde{\gamma}^{}_1 $, coincides with the unretarded contribution 
 to the decay into graphene's longitudinal (charged excitations), previously 
 considered\cite{Swathi09}. 
The other two terms \eqref{second} and \eqref{third}, absent in a non-retarded calculation,
prevail at large distances. The contribution of Eq. \eqref{second} comes from charged
excitations whereas \eqref{second} is due to transverse excitations. Quantitatively, it is the
last term \eqref{third}, which provides the dominant large distance asymptotic behavior.

\subsection{Analytical approximation}
Our analytical approximation consists in setting the functions $\tilde{f}_i$ equal to one. The
approximation $\tilde{f}_i\approx 1$ for $i=1,2$ holds when $x_0 \ll 1$ and $x_0^2
(\frac{\pi^2\alpha_g^2}{4}-1) \ll 1 $, where 
$x_0 = \frac{1}{4 \pi}\frac{v}{c}\frac{\lambda}{z}$, with
{\em graphene's fine structure} constant $\alpha_g = \frac{c}{v} \alpha$.
 For graphene parameters, the latter condition dominates
and can be recast as $ x_0 \lesssim \tfrac{2}{\pi\alpha_g}$ or, equivalently,
$\frac{z}{\lambda}\gtrsim \frac{\alpha}{8} \approx 10^{-3}$, justifying our analytical
approximation. The   approximation   $\tilde{f}_3\approx 1$ applies  for $x_0\ll 1$,  
implying   $\frac{z}{\lambda} \gtrsim \frac{v}{4 \pi c}\approx 10^{-4} $. 

\begin{figure} 
\includegraphics[clip,width=8cm]{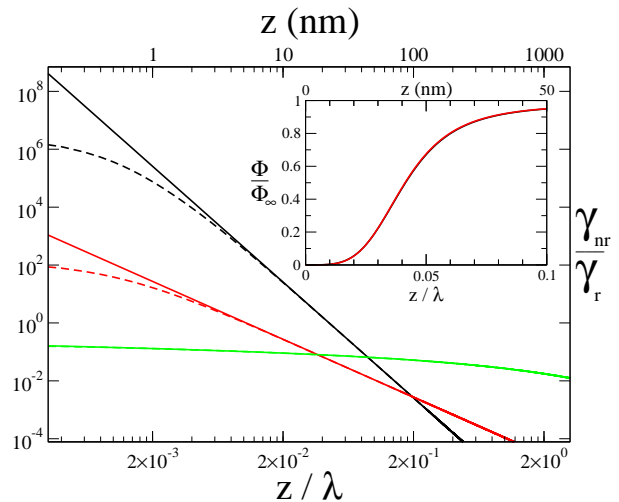} \\ 
\caption{Distance dependence of the analytical approximation for the decay
channels 
($\tilde{\gamma}^{}_i $, solid lines) compared to the exact results 
($\tilde{\gamma}^{}_i\tilde{f}_i $, dashed lines). 
Black solid (dashed) line: $\tilde{\gamma}^{}_1\;(\tilde{\gamma}^{}_1\tilde{f}_1) $.  
Red solid (dashed) line: $\tilde{\gamma}^{}_2\;(\tilde{\gamma}^{}_2\tilde{f}_2) $. 
Green solid (dashed) line: $\tilde{\gamma}^{}_3\;(\tilde{\gamma}^{}_3\tilde{f}_3) $. 
(Exact results correspond to $\lambda = 500 \text{nm}$). Inset: Fluorescence as
a function of distance. Black line: exact result. Red line: analytical approximation.} 
\label{channelsfig}
\end{figure}

The exact and approximate non-radiative 
decays 
are plotted in Fig. \ref{channelsfig}. 
One sees that the approximation only fails in the extreme sub-wavelength regime.
 Furthermore, even though the exact decay rate
saturates for $(z/\lambda)\rightarrow0$ whereas  the approximate one
diverges, this saturation value is so huge that the difference between exact and
approximate results has virtually no impact on $\Phi$ as seen in the inset of Fig. \ref{channelsfig}, where the exact and approximate curves are indistinguishable. Fig. \ref{channelsfig} confirms that retardation channels, $\gamma^{}_{2,3}$, control the decay at large distances, leading to a $z^{-2}$ behavior dominated by graphene's transverse excitations of Eq. \eqref{third}.

\subsection{Fundamental ruler} Notice that only  $\alpha$,  $z$ and $\lambda$ appear in expressions
\eqref{items}, without any reference to graphene's properties. This implies
that a measurement of the fluorescence quenching
amounts to a measurement of the distance $z$, in
terms of the light's wavelength $\lambda$, and the fine structure constant
$\alpha$. In others words, it provides us with a {\em fundamental distance
ruler}.

In general, one would expect graphene's properties to drop out from
dimensionless optical properties involving graphene's excitation
within the light-cone, such as the absorption. There, $\omega \gg vq$,
and graphene's response becomes local, leaving $\alpha$ as the sole
coupling scale. But a cursory application of this reasoning to our
case would justify expressions like Eqs. \eqref{items} only for
distances $z\gtrsim\lambda$, where graphene's light-cone excitations
dominate. Surprisingly, the analytical expression apply for virtually
arbitrary short distances, implying graphene's excitations deep into
the evanescent region where the $q$-dependence does not seem obviously
negligible. A further surprise is the enormous efficacy of graphene as
a fluorescence quencher, particularly in view of the nominally weak
coupling with the field, set by $\alpha$. Both facts, {\em range and
  strength}, can be explained if the naive nominal range for the
expected disappearance of graphene's Fermi velocity $v$ in the
expressions $z\gtrsim\lambda$, can be extended to much shorter
distances $z \ll \lambda$. Our approximations show that this is indeed
the case. The blow up of $(\lambda/z)^4 $ in the first term of Eq.
\eqref{gammaanal} when $(z/\lambda)\rightarrow 0$,
%
more than compensates the
overall factor $\alpha$, leading to strong quenching, as experimentally
observed\cite{Treossi09,Kim10,Chen10}.

\subsection{Finite doping}  
Graphene's fluorescence quenching in doped graphene due to
graphene's longitudinal coupling to the light field has first been analyzed in Ref.
\cite{Velizhanin11} within the unretarded approximation. For large frequencies,
$\hbar\omega\gsim2E_F$, the fluorescence yield of undoped graphene is obtained. For
frequencies $\hbar\omega\lsim2E_F$, the quenching behavior is dominated by
longitudinal plasmon excitations, leading to an characteristic exponential decay
with distance. 

Here, we extend the discussion by also analyzing the 
{\em transverse plasmon} excitations.\cite{Ziegler07}
 These exist in the range 
 $1.667<\Omega\lsim2$ with $\Omega=\omega/(vk_F)$, 
 leading to a decay rate, $\gamma^{}_{t} $, which dominates at long-distances:
\begin{align}
\hbar\gamma^{}_{t}=\frac{\omega^2}{c^2}p_{\shortparallel}^2\frac{q_p^\prime}{4\epsilon_0}e^{-2q_p^\prime z}\frac{1}{1-q_p^\prime\frac{d}{dq'}\ln\chi_t|_{q'=q_p^\prime}}
,\end{align}
where $ q_p^\prime=\sqrt{q_p^2-(\omega/c)^2}$, with plasmon momentum $q_p$. Approximating graphene's transverse response by the long-wavelength limit, we obtain the following analytical expression:
\begin{align}
\frac{\gamma^{}_{t}}{\gamma^{}_0}=\frac{3\pi}{2}\frac{p_\shortparallel^2}{p^2}\frac{\alpha}{\Omega}f(\Omega)e^{-z/z_0}\;
\end{align}
with $z_0^{-1}=4\alpha(v_F/c)k_Ff(\Omega)$ and
\begin{align}
f(\Omega)=\frac{\Omega}{4}\ln\left|\frac{2+\Omega}{2-\Omega}\right|-1
.\end{align}

\begin{figure} 
\includegraphics[clip,width=8cm]{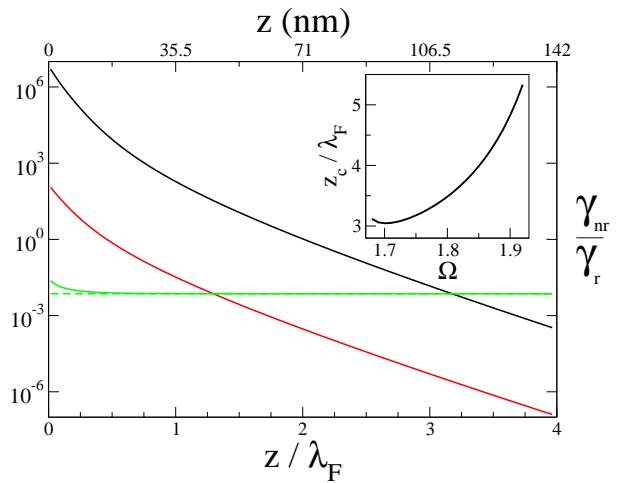} \\ 
\caption{Distance dependence for the decay
channels ${\gamma}^{}_i $ for doped graphene for $\Omega=1.75$. Black solid line: ${\gamma}^{}_1$.  
Red solid line: ${\gamma}^{}_2$. 
Green solid line: ${\gamma}^{}_3$. Green dashed line: ${\gamma}^{}_t$.
Inset: crossover length as a function of frequency. (Numerical values correspond to $\lambda = 500 \text{nm}$ and electronic density $n=10^{12}\text{cm}^{-2}$)} 
\label{dopedchannelsfig}
\end{figure}

In Fig. \ref{dopedchannelsfig}, the distance dependence of the various
decay channels is shown for $\Omega=1.75$ as obtained numerically. The
response is controlled by the singular features in the dispersion
relation, leading to an exponential distance law.  The transverse
decay channel ${\gamma}^{}_3$ is almost entirely due to the transverse
plasmon mode ${\gamma}^{}_t$, and dominates the long-distance behavior
beyond a crossover length, $z_c$, whose frequency dependence is shown
in the inset of Fig. \ref{dopedchannelsfig}.
We finally note, that the extremely slow decay rate of ${\gamma}^{}_t $ as well as the large distance required for the onset of the power law in Eq. \eqref{third} can be linked to the condition $ 1 -  d_t \chi^{}_t\approx 0$.

\section{Summary}
\label{summary} 
We first studied the fluorescence
quenching efficacy of undoped graphene as function of distance,
including transverse decay channels, retardation and graphene-field
coupling to all orders. For shorter distances, we confirm the validity
of previous lowest-order, unretarded results, albeit with
modifications in range and saturation value. 
Retardation changes the distance law
to $(\lambda/z)^2 $, with both longitudinal (charged) and transverse
(chargeless) graphene's excitation contributing to it, the latter
dominating in the truly large distance asymptotic regime. 

A compact, virtually exact analytical expression has been obtained for
the zero-doping fluorescence yield for all distances. It involves only $\alpha$
and $z/\lambda$, and, therefore, endows graphene's fluorescence
quenching measurements with the unique status of a fundamental ruler.

We also showed that measurements of the fluorescence quenching
efficacy of doped graphene at appropriate frequencies and large
distances give direct evidence of the existence of transverse plasmons
unique to two-band materials like graphene.

Finally, we note that our results might also be important because 
non-radiative decay of fluorescent materials amounts to electron-hole
(carrier) generation in graphene. Direct (propagating field)
photo-generation of carriers is inefficient in graphene
$(\text{absorption} \approx \pi \alpha)$, whereas indirect
photo-generation by the evanescent field (that is, fluorescence
quenching) can be very effective, ultimately controlled by the dye's
absorption and quantum yield. This suggests new ways of enhancing
graphene's photo-responsivity\cite{Milan11,Thongrattanasiri11,Echtermeyer11}.
 
\section{Acknowledgments} This work has been supported by FCT under grant PTDC/FIS/101434/2008 and MIC under grant FIS2010-21883-C02-02. 
\appendix
\section{Explicit form of decay rates}
\label{Appendix}
Here we outline the derivation of the non-radiative decay
channels at zero doping
\begin{equation}\label{agammarnr}
\hbar \gamma^{}_{nr} = -2 \omega^2 p_{\alpha}^{*}  \; 
\{\frac{1}{(2 \pi)^2}\int_{q>\omega/c} d^2q \,
                   \text{Im}\mathcal{D}^{\alpha \beta}(z,z)\}             
                \;p_{\beta}
.\end{equation}

As the non-radiative decay rate only
exists in the presence of graphene, we classify its
contributions according to the nature of graphene's excitations:
longitudinal (charged) and transverse. Longitudinal $(l)$ excitations couple to both
in-plane $(p_{\shortparallel})$ and out-of-plane $(p_z)$ components of the
dipole matrix elements. Graphene's transverse $(t)$ excitations only couple to the 
in-plane $(p_{\shortparallel})$ dipole matrix element. Therefore, we write for
the non-radiative decay channels

\begin{equation}\label{channels}
\gamma^{}_{nr} = \gamma^{}_{l,\shortparallel} +\gamma^{}_{l,z}+\gamma^{}_{t,\shortparallel}\;.\end{equation}

Let us consider  $\gamma^{}_{l,\shortparallel}$ first.
Selecting the longitudinal part of Eq. \eqref{chiij} in  Eq. \eqref{Dgraphene}, 
it is  straightforward to show that Eq. \eqref{agammarnr} leads to
\begin{equation}\label{gamma1}
\hbar \gamma^{}_{l,\shortparallel}= -2 \omega^2 \frac{p_{\shortparallel}^2}{4 \pi}  \; 
\int_{0}^{\infty} q'dq' {\rm e}^{-2 q'z} \,
                  \text{Im} \frac{  d_l^2 \chi^{}_l}{1 -  d_l \chi^{}_l}            
,\end{equation} 
where we have traded $q$ for $q'$ in the integration. 
Notice that, although the exact (i.e. retarded) photon
propagators are used, the only manifestation of light's velocity $c$ in $
\gamma^{}_{l,\shortparallel} $ would be  the substitution: 
$ v^{-2}\rightarrow\tilde{v}^{-2}=v^{-2} - c^{-2}$,  
within the square root of $\chi^{}_l$ at zero doping.
  The quantitative irrelevance of
this replacement makes the instantaneous approximation for
$\gamma^{}_{l,\shortparallel} $ correct. The integration in Eq. (\ref{gamma1}) 
then leads to the first term of $\gamma_{nr}/\gamma_r$ given in the main text,  with the dimensionless factor
$\tilde{f}_1$   given by
\begin{equation}\label{ftilde1alt2}
 \tilde{f}_1= \frac{1}{3! x_0^4}
\int_0^1 dx\,  \frac{x^3 {\rm e}^{-x/x_o}\sqrt{1-x^2}}{1 -x^2 + 
\frac{\pi^2 \alpha_g^2 }{4} x^2}          
,\end{equation} 
where $x_0 = \frac{1}{4 \pi}\frac{\tilde{v}}{c}\frac{\lambda}{z}$, with 
{\em graphene's fine structure} constant $\alpha_g = \frac{c}{v} \alpha$.
 The approximation $\tilde{f}_1\approx 1$ holds when $x_0 \ll 1$ and 
 $x_0^2 (\frac{\pi^2\alpha_g^2}{4}-1) \ll 1 $. For graphene parameters, the
 latter condition dominates and can be recast as
 $ x_0 \lesssim \tfrac{2}{\pi\alpha_g}$ or, equivalently,
 $\frac{z}{\lambda}\gtrsim \frac{\alpha}{8} \approx 10^{-3}$, justifying our approximation $\tilde{f}_1\approx1$.

Now, we consider the non-radiative decay channel coupling graphene's 
longitudinal response 
with an out-of-plane dipole.
Using Eqs. \eqref{D0izexplicit} and  \eqref{Dgraphene}, Eq. \eqref{agammarnr}
leads to  
\begin{equation}\label{gamma2}
\hbar \gamma^{}_{l,z}= -2 \omega^2 \frac{p_{z}^2}{2 \pi}  \; 
\int q'dq' {\rm e}^{-2 q'z} \, \left(1 + \frac{\omega^2}{c^2 q'^2}\right)
                  \text{Im} \frac{  d_l^2 \chi^{}_l}{1 -  d_l \chi^{}_l}            
.\end{equation} 
 The integration can be rewritten as the two first terms of $\gamma_{nr}/\gamma_r$ given in the main text, where the new dimensionless factor $\tilde{f}_2$, corresponding to the second
 term in the sum of Eq. \eqref{gamma2}, appears. It is given by
\begin{equation}\label{ftilde2alt2}
 \tilde{f}_2= \frac{1}{x_0^2}
            \int_0^1 dx\,  
	    \frac{x {\rm e}^{-x/x_o}\sqrt{1-x^2}}{1 -x^2 + \frac{\pi^2 \alpha_g^2 }{4} x^2}              
.\end{equation} 
The approximation  $\tilde{f}_2\approx 1 $ has the same range of validity as
that of $\tilde{f}_1 $.

Finally, we consider  graphene's transverse excitation channels. Using the
transverse components of Eqs. \eqref{D0explicit} and \eqref{chiij} in
\eqref{Dgraphene}, Eq. \eqref{agammarnr} leads to 
\begin{equation}\label{gamma3}
\hbar \gamma^{}_{t,\shortparallel} = -2 \omega^2   \frac{p_{\shortparallel}^2}{4 \pi}\; 
              \int q'dq' {\rm e}^{-2 q'z} \, 
              \text{Im} \frac{  d_t^2 \chi^{}_t}{1 -  d_t \chi^{}_t}            
,\end{equation} 
an integration that can be rewritten as third term of $\gamma_{nr}/\gamma_r$ given in the main text, with the
corresponding dimensionless parameter $\tilde{f}_3$ given by
\begin{equation}\label{ftilde3}
 \tilde{f}_3= \frac{1}{{\rm g}(2 \pi^2 \alpha \frac{z}{\lambda})}
            \int_0^1 dx\,  
	    \frac{x {\rm e}^{-x/x_o}\sqrt{1-x^2}}
	    {x^2 +  (\frac{\pi v}{2 c}\alpha)^2 (1 -x^2)}              
,\end{equation} 
with the function ${\rm g}(a)$ defined\cite{Gradshteyn80} as 
\begin{equation}\label{g(a)}
 {\rm g}(a)= \int_{0}^{\infty} dx \, \frac{x {\rm e}^{-x}}{x^2 + a^2} = -{\rm ci}(a)
 \cos(a) - {\rm si}(a)\sin(a)
.\end{equation} 
The   approximation   $\tilde{f}_3\approx 1$ applies  for $x_0\ll
1$,   implying   $\frac{z}{\lambda} \gtrsim \frac{v}{4 \pi c}\approx 10^{-4}
$, as stated in the main text.

 The function ${\rm g}(a)$ exhibits the following asymptotics\cite{DLMF1}:
\begin{equation}\label{g(a)aprox}
\begin{split}
 {\rm g}(a\gg1)&= \frac{1}{a^2} ( 1 - \frac{3!}{a^2} + \frac{5!}{a^4} - \cdots )\\ 
 {\rm g}(a\ll1)&= -\log({\rm e}^{\cal C} a)
,\end{split} 
\end{equation} 
with Euler constant $\cal C$. The first limit in Eq. \eqref{g(a)aprox} for
$a=2 \pi^2 \alpha \frac{z}{\lambda}\gg 1$ explains the large distance
asymptotic behavior of $\tilde{\gamma}^{}_3 $ as discussed in the main text.
\bibliography{suscarga,fluor}
\end{document}